\documentclass[11pt, oneside]{article}
\usepackage{geometry}
\usepackage{graphicx}
\usepackage{amssymb}
\usepackage{amsmath,amsfonts,amsthm,mathtools}
\usepackage{booktabs}
\usepackage[round]{natbib}
\usepackage{url}
\usepackage{setspace}
\usepackage{authblk}

\title{The Prosecutor's Fallacy and Expert Testimony: \\ A Modern Take Using Likelihood Ratios}

 \author[1,2]{Maria Cuellar}
 
 \affil[1]{Department of Criminology, University of Pennsylvania, 3718 Locust Walk, Philadelphia, PA, 19104, United States}
 
 \affil[2]{Department of Statistics and Data Science, Wharton School, University of Pennsylvania, Walnut Street, Philadelphia, PA 19104, United States}

\date{January 30, 2025}

\begin{document}
\maketitle

\begin{abstract}
Forensic examiners and attorneys need to know how to express evidence in favor or against a prosecutor's hypothesis in a way that avoids the prosecutor's fallacy and follows the modern reporting standards for forensic evidence. This article delves into the inherent conflict between legal and scientific principles, exacerbated by the prevalence of alternative facts in contemporary discourse. Courts grapple with contradictory expert testimonies, leading to a surge in erroneous rulings based on flawed amicus briefs and testimonies, notably the persistent prosecutor's fallacy. The piece underscores the necessity for legal practitioners to navigate this fallacy within the modern forensic science framework, emphasizing the importance of reporting likelihood ratios (LRs) over posterior probabilities. Recognizing the challenge of lay comprehension of LRs, the article calls for updated recommendations to mitigate the prosecutor's fallacy. Its contribution lies in providing a detailed analysis of the fallacy using LRs and advocating for a sound interpretation of evidence. Illustrated through a modified real case, this article serves as a valuable guide for legal professionals, offering insights into avoiding fallacious reasoning in forensic evidence assessment.
\end{abstract}

\section{Introduction}

There is a fundamental tension between law and science due to their divergent principles. Legal decisions seek finality and consistency through precedent, while scientific results change with new evidence. In the current era of ``alternative facts,'' courts grapple with contradictory and irreconcilable expert testimonies \citep{balkoaltfacts}, leading to a large and growing number of court rulings that have relied on expert testimony and assertions in amicus briefs that are patently incorrect \citep{balkoaltfacts, liptakfacts}.

This is especially important today because attorneys, both in the defense and prosecution, need to avoid incorrect statements within the modern framework of forensic science reporting. Recently, researchers have argued that experts should not comment on the posterior probability of guilt or innocence of the defendant, nor on the posterior probability that two items have the same source or a different source \citep{thompson2018after}. Posterior probabilities require experts to ``overreach'', i.e., to go beyond their forensic expertise and to invade the province of the judge and jury by opining about priors and on the weight of non-forensic evidence. Instead, in order to stay within their expertise, experts should limit themselves to commenting about the probability of observing what they see in the evidence under the relevant hypotheses. Experts can do this is by reporting likelihood ratios, instead of a posterior probability or posterior odds, to provide a measure of uncertainty for a conclusion about the evidence  \citep{thompson2018after}. Even though likelihood ratios can be difficult for lay persons to understand \citep{thompson2015lay}, they are recommended by authorities \citep{champod2016enfsi} as the way experts should be reporting conclusions, and they avoid having experts report posterior probabilities. 

One common error in expert report testimony is called the ``prosecutor's fallacy'', a logical error of believing that the chance of a rare event happening is the same as the chance of a suspect's innocence. Even though it was identified as far back as 1987 \citep{thompson1987interpretation}, both prosecution and defense continue to suffer from this error \citep{leung2002prosecutor, kilberger2023something, de2008guilt}, and is even found in other disciplines, including medicine and epidemiology  \citep{greenland2011null, westreich2014epidemiology}. One way that experts have been advised to avoid the prosecutor's fallacy when they give reports requires directly reporting about the defendant's guilt in light of the evidence \citep{cebm}. However, this implies reporting posterior probabilities. 

Recommendations for how to avoid the prosecutor's fallacy are outdated because they involve reporting posterior probabilities. However, the literature is lacking guidance on how to avoid the prosecutor's fallacy while reporting conclusions as likelihood ratios, i.e., without posterior probabilities. Avoiding the prosecutor's fallacy while reporting a likelihood ratio poses a challenge to expert witnesses and attorneys seeking to make correct statements. 

The contribution of this article is to provide a description of the prosecutor's fallacy using likelihood ratios and to present a correct interpretation of the evidence. This article provides an example from a recent real case, for which the details have been modified for privacy purposes, in which the prosecution suffered from the prosecutor's fallacy. Clear guidance on how to report conclusions within the modern framework of likelihood ratios, while avoiding the prosecutor's fallacy, can ease the tension between science and the law because it can prevent legal actors from making fallacious arguments. This type of guidance can help bring transparency to how statistics should be used in the law, such that attorneys can understand and feel at ease with statistical claims.


\section{Background}

\subsection{The Prosecutor's Fallacy}\label{sec:backgroundprosfallacy}

The prosecutor's fallacy is a fallacy of statistical reasoning, usually committed by the prosecution, to argue for a defendant's guilt during a criminal trial. Although it is named after prosecutors, it has also been used by defense attorneys arguing for the defendant's innocence. A fallacy is the use of invalid or otherwise faulty reasoning in the construction of an argument. The term ``prosecutor's fallacy'' was first used by \cite{thompson1987interpretation}. Here, I present two examples to illustrate the fallacy, a toy example and one from the real world. In the next section, I formalize how some have recommended repairing the fallacy, and how I suggest it be done instead, using the likelihood ratio.

First, I consider a toy example. Suppose that in a city with population of two million, a man stole a woman's purse. An eyewitness claimes that the criminal had red hair and brown eyes. A database (e.g., DMV) shows that there are two hundred people living in the city who have red hair and brown eyes. The police find that a man with red hair and brown eyes lived near the crime scene, and they charge him with the crime. The prosecuting attorney then argues that since the defendant's hair and eye color is consistent with the criminal's, the defendant is guilty of the crime. However, this reasoning is incorrect.

Second, an infamous example of the prosecutor's fallacy from the real world is the case of Sally Clark. In the late 1990's, Sally Clark's two-month-old previously healthy infant died. Almost two years later, Clark's second child died in very similar circumstances, also at two months old. Although there was no other evidence pointing to this, Clark was accused of murdering her two children. A statistical expert, Sir Roy Meadows, argued that it is very rare that a child die of SIDS (a chance of 1/8,500 in the UK at the time), and it is even more rare to have two children die of SIDS. He argued that the chance of the two children dying of SIDS was 1/8,500 squared or 1 in 73 million chance that the mother did not kill them. Meadows assumes that the two deaths were independent \citep{rsssallyclark}. But this reasoning is incorrect \citep{csafepf}.


\subsection{The Likelihood Ratio}

There are various ways of characterizing the strength of a forensic scientist's source conclusions. There is no consensus yet on which reporting method is best, in part because it is not clear how the various possible statements are understood by the target audience. \cite{thompson2018perceived} examined how lay people evaluate the strength of different reporting statements relative to one another. Types of reporting statements include categorical conclusions (CC), random match probabilities (RMPs), likelihood of observed similarity (LOS), source probability statements (SP), and strength of support statements (SOS). 

Likelihood ratios (LRs) are used by experts to describe their assessments of the strength of forensic evidence \citep{aitken2020statistics}. The use of LRs is recommended by the European Network of Forensic Science Institutes (ENFSI) and the UK Royal Statistical Society. Scientists in Europe, New Zealand, and parts of Australia also have adopted this approach \citep{biedermann2017development}, and scientists in the United States have used this as well, although primarily for DNA \citep{butler2009fundamentals}. 

Dennis \cite{lindley} laid the groundwork for a statistical treatment of evaluating the weight of evidence in forensic science. His framework, updated in \cite{lindley2013understanding}, and later used by researchers (e.g., \cite{lund2022bayesian}), was based on a subjective Bayesian modeling of the problem. Using the odds form of Bayes' rule, it said that the decision-maker (DM), which is the judge or jury, should learn a forensic expert's (Expert) testimony as a likelihood ratio, to update their beliefs about whether the defense or prosecution's hypothesis is more likely,
\begin{equation}
\text{Posterior Odds}_{\text{DM}} = \text{Likelihood ratio}_{\text{Expert}} \times \text{Prior Odds}_{\text{DM}}.
\end{equation}

More specifically, suppose the forensic expert is responsible for determining whether two pieces of evidence, $x$ and $y$ (denoted jointly by $E$), were produced by the same source. The expert's task is comparing the probability that samples $x$ and $y$ have a common source, an event denoted by $H_p$ for the prosecution's hypothesis, given $x$, $y$, and the information $I$, i.e., $P(H_p|x,y,I)$, to the probability that $x$ and $y$ have a different source ($H_d$) for the defense hypothesis, given the same information, $P(H_d|x,y,I)$. The case information $I$ should be relevant to the forensic task in order to avoid contextual bias \citep{cuellar2022probabilistic}. For the rest of this article, we omit the information $I$ from the equations because we focus on the information learned directly from the evidence. After receiving the evidence and other case information, the expert should estimate LR, and should communicate it to the decision-maker so they can update their beliefs concerning the event $H_p$. This task can be written by using the odds form of Bayes' rule,
\begin{equation}
\label{eq:bayesruleanalyst}
\underbrace{\frac{P(H_p|E)}{P(H_d|E)}}_{\text{Posterior odds}_{\text{DM}}} =  \underbrace{\frac{P(E|H_p)}{P(E|H_d)}}_{\text{Likelihood ratio}_{\text{Expert}}} \times \underbrace{\frac{P(H_p)}{P(H_d)}}_{\text{Prior odds}_{\text{DM}}}.
\end{equation}
The prior odds is the decision-maker's beliefs about $H_p$ and $H_d$ prior to seeing the evidence, the likelihood ratio is the quantity at which the expert arrives through the analysis of the evidence, and the posterior odds is the final quantity at which the decision-maker will arrive after rationally incorporating all available information. 

How to estimate the LR is a controversial question, and whether it is even considered a likelihood ratio or a Bayes' factor is a subtle issue that requires careful analysis \cite{ommen2021problem}. In this article, I provide an example for how the LR could be calculated in a real legal case, and I point to where the controversies arise in the calculation.

To make it easier for lay people to understand the meaning of a LR, the European Network of Forensic Science Institutes (ENFSI) has provided a verbal expression of probative strength for values of the LR \citep{ensfireport}. For example, if a forensic scientist concludes that the comparison is 5,000 times more likely if the samples have a common source rather than a different source, the scientist should report that the comparison provides ``strong'' support for the conclusion that the items have a common source. If LR = 500 this is ``moderately strong'' support, 50 is ``moderate'' support, etc.

There is less agreement about how the expert should provide the decision-maker with a LR. The expert could state their LR and leave it up to the trier of fact to decide what to do with it, or the expert could tell decision-makers directly to multiply their own prior odds by the expert's LR. Furthermore, how the decision-maker should incorporate multiple likelihood ratios from a variety of experts, and whether they should treat these as independent, is an interesting question, but outside the scope of this article. \cite{cuellar2022probabilistic} has a description of how the trier of fact incorporates information from several experts in Section 5. More recent work by \cite{lund2024influence} has argued, interestingly, that Bayesian reasoning does not support using someone else's LR in place of one's own, and have suggested instead that the expert's role is to provide information that empowers a lay recipient to formulate their own LR informally for the evidence considered by the expert.

For this article, we use the framework from Equation \ref{eq:bayesruleanalyst}, recommending that the expert's LR is directly used by the DM to update their beliefs. But, further work should focus on how this formulation would change if the expert's LR is given to the DM and then the DM informally formulates their own LR, incorporating information about how trustworthy the expert is, among other pieces of information.

\section{Motivating Legal Case}

\subsection{Setup}

These are the facts of the case:
\begin{enumerate}
\item A man (Mr. A) was murdered. When his body was found, he was wearing a jacket that originally had two zippers. One of the pull tabs from the zippers was missing. 
\item A year after the homicide, the police found one zipper pull tab in the defendant's (Mr. B's) backyard. 
\item An expert stated that the zipper pull tab found in the backyard was consistent with the zippers from the jacket found on the body. 
\item Another expert who works at a local store (call it Store C), stated that the jacket the man was wearing was sold by Store C two years prior to the homicide. 
\item The prosecution hired the worker at Store C manufacturing who was in charge of ordering zippers for this jacket as well as other clothing items at Store C. The worker stated that he had ordered the manufacturing of 5,000,000 zippers for manufacturing in the entire state the year prior to the homicide. 
\item The prosecution argued that the zipper pull tab from Mr. B's backyard must belong to Mr. A, and that the fact that this pull tab was found in the suspect's property suggests that the defendant killed Mr. A.
\end{enumerate}

An expert in statistics was asked, does the fact that the zipper pull tab from the suspect's (Mr. B's) house is consistent with the one at the crime scene imply that Mr. B committed the crime? And if so, how strong is the weight of the evidence?

In a criminal proceeding, there are two hypotheses in question. $H_d$ is the defense hypothesis that no misconduct has taken place, and $H_p$ is the prosecution hypothesis that misconduct has taken place. Let $E$ be the evidence that is found. In this case,
\begin{itemize}
\item $H_p$: Zipper pull tab in defendant's property belongs to the victim's jacket.
\item $H_d$: Zipper pull tab in defendant's property does not belong to the victim's jacket.
\item $E$: Zipper pull tab in defendant's property is consistent with the victim's jacket.
\item $\neg E$: Zipper pull tab in defendant's property is not consistent with the victim's jacket.
\end{itemize}

With the evidence provided in the facts of the case, Table \ref{tab:evidence} is obtained. However, the number of zipper pull tabs that are not consistent with the pull tab from the crime scene, $P(H_d)$, is missing, as well as all the values that require this number for calculation. How could one know how many zipper pull tabs different from the ones on Mr. A's jacket were in the state at this time? Any guess is bound to be incorrect. This is precisely why these calculations are controversial. Someone needs to speculate about this number to estimate a likelihood ratio.

\begin{table}[h] \centering 
\begin{tabular}{c|cc|c} 
\toprule
	& $H_p$  & $H_d$        & Total       \\
	\midrule
	 $E$ & 1     & 5,000,000   & 5,000,001   \\
	$\neg E$ & 0     & ??? & ??? \\
	\midrule
	Total & 1     & ??? & ???\\
\bottomrule
\end{tabular}
\caption{Evidence from the case in table form. The four cells with question marks show the values that cannot be known from the facts presented in the case.}
\label{tab:evidence}
\end{table}

If the statistical expert guesses the number, this is outside that expert's area of expertise. Statisticians do not specialize in zipper pull tabs and their frequency in a state at a given time. However, to arrive at any estimate of a likelihood ratio, the statistician needs to have this number. One way to do it is to guess. However, researchers warn against this when they say, ``If the examiner does not know enough to assess the relevant probabilities, then the examiner does not know enough to evaluate the strength of the forensic evidence --- and hence nothing the examiner says about the value of the evidence should be trusted.'' \citep{thompson2018after} Despite this valid warning, let us guess the number: The year of the murder in the state, there were 100 times more zipper pull tabs that were not consistent with the zipper pull tab at the crime scene than pull tabs that were consistent with it. With this, we can finish populating Table \ref{tab:evidencefull}. Armed with this table, we can precisely state the fallacy in the prosecutor's argument. I return to this guess in Section \ref{sec:guess}.


\begin{table}[h] \centering 
\begin{tabular}{c|cc|c} 
\toprule
	& $H_p$  & $H_d$        & Total       \\
	\midrule
	 $E$ & 1     & 5,000,000   & 5,000,001   \\
	$\neg E$ & 0     & 500,000,000 & 500,000,000 \\
	\midrule
	Total & 1     & 505,000,000 & 505,000,001\\
\bottomrule
\end{tabular}
\caption{Evidence from the case in table form, with the guess that the year of the murder in that state, there were 500 million zipper pull tabs that were not consistent with the one at the crime scene.}
\label{tab:evidencefull}
\end{table}

\section{Analysis}

\subsection{The Previous Way}

In the case in question, the fallacy happens when the prosecutor says, ``What are the chances that the zipper pull tabs would be consistent? There are lots of zipper pull tabs out there! It is very unlikely that we got a match just by chance.'' In other words, according to the evidence from Table \ref{tab:evidence}, the fallacy is when the prosecutor presents the number 
\begin{equation}
P(E|H_d) = \frac{5,000,000}{505,000,000} = 0.01, 
\end{equation} 
as the probability that the defendant is innocent, given the evidence. It is so small, that it must be that the defendant is guilty. 

The old way of correcting this was to recommend that the prosecutor say instead, ``Actually, there are lots of zipper pull tabs that would match this one, because it is a very common type of zipper pull tab. So, it is very likely that the pull tab found at the defendant's house is not from the victim's jacket.'' In other words, the prosecutor should calculate 
\begin{equation}
P(H_d | E) = \frac{5,000,000}{5,000,001} = 0.99, 
\end{equation}
instead as the probability that the defendant is innocent, given the evidence. Thus, the probability the defendant is guilty is very small. 

The prosecutor's fallacy used to be presented as the error that occurs when someone, traditionally the prosecution, assumes incorrectly that $P(E | H_d)$ being small entails that the posterior probability $P(H_d | E)$ is small.\footnote{Note that sometimes the events are not $H_d$ vs.~$H_p$, but the fact that the defendant is innocent vs. guilty. We present $H_d$ vs.~$H_p$ to be more general, and because we follow the guidance here that the expert should never opine about the defendant't guilt, only about the evidence.} This implies that the prosecution hypothesis given the evidence is high. \cite{cebm} In other words, the prosecutor's fallacy incorrectly assumes that,

\vspace{.2in}
\noindent\fbox{%
    \parbox{\textwidth}{%
Prosecutor's fallacy (old way): 
\begin{equation}
P(E | H_d) \text{ is small} \implies P(H_d | E) \text{ is small} \implies  1-P(H_d|E) = P(H_p | E) \text{ is large}.
\end{equation}
    }%
}
\vspace{.2in}

\noindent This reasoning is fallacious because it relies on the false equivalence,
\begin{equation}
P(E | H_d) = P(H_d | E), 
\label{eq:flippingconditional}
\end{equation}
which is not true in general. For this reason, sometimes the error in the prosecutor's fallacy is called ``flipping the conditional'' or ``transposition of the conditional''. 

Translating these conditional probabilities to words is not particularly helpful. In fact, it confuses the point and makes it difficult to see why the statement is incorrect. Putting Equation \eqref{eq:flippingconditional} into words suggests why this fallacy might be happening in the first place. It is confusing these two statements, which are equivalent in words, but not in conditional probabilities.
\begin{itemize}
\item ``The probability that the evidence is found given that no misconduct has taken place.''
\item ``The probability that no misconduct has taken place given that the evidence is found.'' 
\end{itemize}

Some (e.g., \cite{cebm}) have suggested that the fallacy can be prevented by having the expert state $P( H_d | E )$ instead of $P( E | H_d)$ as the probability of interest, or the chance that the defense is correct given the evidence. However, this requires making a statement about posterior probabilities, about whether the expert thinks that the defendant is innocent or guilty of the crime. That is outside the responsibility of the expert, and it is a decision that should be made by the trier of fact. The new recommendations are that the expert should not opine about posterior probabilities, such as $P(H_d | E)$ \citep{thompson2018after}. So, what should an expert do instead? Reporting the likelihood ratio requires a different approach.

\subsection{The New Way}

When an expert reports the likelihood ratio, the prosecutor's fallacy assumes that the likelihood ratio equals the posterior odds, or, equivalently, that the prior odds equal one. That is, since 
\begin{equation}
LR = \frac{P(E|H_p)}{P(E|H_d)} = \frac{1}{5,000,000/505,000,000} = 101,
\end{equation}
the fallacy assumes incorrectly that the posterior odds are also 101.  But, clearly the prior odds are not equal to one. Instead, 
\begin{equation}
\text{Prior odds} = \frac{P(H_p)}{P(H_d)} = \frac{1}{505,000,000} = 0.000000002.
\end{equation}
So, combining these using Bayes' rule, the correct posterior odds is minuscule,
\begin{equation}
\text{Posterior odds} = LR \times \text{Prior odds} = 101 \times 0.000000002 = 0.000000202.
\label{eq:posterioroddscorrect}
\end{equation}

In other words, the prosecutor's fallacy is incorrectly assuming that,

\vspace{.2in}
\noindent\fbox{%
    \parbox{\textwidth}{%
Prosecutor's fallacy (new way): 
\begin{equation}
LR \text{ is large} \implies \text{Posterior odds} = \frac{P(H_p | E)}{P(H_d | E)} \text{ is large}.
\end{equation}
    }%
}
\vspace{.2in}

\noindent This is fallacious because it relies on the false equivalence,
\begin{equation}
\frac{P(H_p)}{P(H_d)} = 1,
\end{equation}
which is not true in general. So, when a forensic expert reports a likelihood ratio, and this is interpreted by the prosecutor (or the decision maker) as the posterior odds, that is the prosecutor's fallacy. 

To avoid the prosecutor's fallacy, the expert should report the LR, and the decision-maker (trier of fact) should multiply that by their prior odds. This might happen informally with verbal equivalents.

\section{Guessing outside the statistician's area of expertise} \label{sec:guess}

Let us return to the guess made in Table \ref{tab:evidencefull}. Note that the values of the LR, prior odds, and thus posterior odds, might be sensitive to the estimate of how many zipper pull tabs that \textit{do not} match the one at the crime scene there exist. This highlights one of the reasons the likelihood ratio is controversial: Who decides what that number should be? Should it include all the zipper pull tabs in the city, or county, or state, manufactured that year? Or in the previous decade, since most of those might still be around? What if individuals from other states bring zipper pull tabs into that state? 

Someone needs to make an educated guess of this number to estimate LR. So far, the forensic statistics community recommends that this task be left up to someone other than the statistical expert, probably the trier of fact. But how can a lay juror, or even a judge, have an educated guess about the number of zipper pull tabs? And even if an expert were to suggest a number, these kinds of quantities are difficult to estimate correctly. Nevertheless, this kind of analysis is similar if the number of pull tabs not matching the evidence is greater than the number of pull tabs matching the evidence. What if it is less?

One could make a guess, and then test the sensitivity of the final result to that guess. Let us vary the guess for the number of zipper pull tabs that are not consistent with the one found at the crime scene. Table \ref{tab:guesses} shows the results. 

\begin{table}[h] \centering 
\begin{tabular}{cccc}
\toprule
	Guess & LR & Prior odds & Posterior odds     \\
	\midrule
0 & 1 & 2.00$\times 10^{-7}$ & 2.00$\times 10^{-7}$ \\ 
10,000,000 & 3 & 6.67$\times 10^{-8}$ & 2.00$\times 10^{-7}$ \\ 
100,000,000 & 21 & 9.52$\times 10^{-9}$ & 2.00$\times 10^{-7}$ \\ 
500,000,000 & 101 & 1.98$\times 10^{-9}$ & 2.00$\times 10^{-7}$ \\ 
1,000,000,000 & 201 & 9.95$\times 10^{-10}$ & 2.00$\times 10^{-7}$ \\ 
1,000,000,000,000 & 200,000 & 1.00$\times 10^{-12}$ & 2.00$\times 10^{-7}$ \\ 
\bottomrule
\end{tabular}
\caption{Sensitivity of the posterior odds to different guesses of how many zipper pull tabs there were that year in that state that are not consistent with that from the crime scene.}
\label{tab:guesses}
\end{table}

Regardless of the guess, it is clear that the posterior odds is small even if the LR is large. Note that in this example, the posterior odds stay constant because the guess is used in the calculation of LR and prior odds, so when they are multiplied to obtain the posterior odds the result is the same. For instance, if the expert guesses that there are one trillion zipper pull tabs in the state in that year that are not consistent with the one at the crime scene, the LR is high at 200,000, but the posterior odds is $2\times 10^{-7}$. Since the LR is high, if one assumes incorrectly that the posterior odds is also high, then that reasoning suffers from the prosecutor's fallacy. But without the prosecutor's fallacy, the trier of fact should arrive at a minuscule posterior odds showing that it is much more likely that the defense hypothesis is correct given the evidence, than the prosecution hypothesis is correct given the evidence.

\section{Discussion}

This article presents an example from a real legal case, for which the details have been removed for reasons of privacy, to exemplify how a prosecutor can fall prey to the prosecutor's fallacy. The expert can present posterior odds, but according to the recommendations of the forensic statistics community and governments, the statistical expert should report the likelihood ratio instead. 

In the legal case in question, the likelihood ratio suggests that the evidence is more in agreement with the prosecution than the defense. If one includes the prior odds, however, one realizes that the posterior odds is minute. It is crucial to realize that it is not only the likelihood ratio that matters, but how it multiplies with the prior odds to produce a posterior odds. The statistical expert should report the likelihood ratio, as well as instructions for how the trier of fact should incorporate this number (or its verbal equivalent) to their reasoning, even if this is in a qualitative an informal way. In addition, if the statistician can provide a guess for the unknown values in the facts of the case, they should do so, provided that they also show how sensitive the posterior odds is to the guess.

This article shows how simple guidance can be used by statistical experts to estimate probabilities correctly. When experts are hired, they are often not aware of common fallacies, and they might make mistakes that have been commonly made in the past. This article has the potential to improve the quality of scientific testimony in criminal and civil trials that use statistical evidence, and thus, it can contribute to ease the tension between the law and science.

\section*{Acknowledgments}

Thank you to Alicia Carriquiry, Steve Lund, and Amanda Luby for stimulating and insightful conversations about this topic, as well as for feedback on my manuscript.

\bibliographystyle{plainnat}

\bibliography{prosecutorsfallacy}

\end{document}